# MANIAC Challenge: The Wolf-pack strategy

Cristian Chilipirea, Andreea-Cristina Petre, Ciprian Dobre
Faculty of Automatics and Computer Science
University POLITEHNICA of Bucharest
Bucharest, Romania
{cristian.chilipirea, andreea.petre}@cti.pub.ro, ciprian.dobre@cs.pub.ro

*Abstract*— MANIAC Challenge raises a problem of game theory, different players strategies intertwine and the success of any player is dependent on the actions of all players in the system. A truly fair scenario is when all the strategies are identical, all the nodes co-operate and they all equally share the rewards and risks that come with every transfer. A successful strategy is one that tries to diverge from the equilibrium to maximize its own gains and it manages to do so. We propose the wolf-pack strategy. Unlike standard game-theory based strategies, our strategy does not penalize the nodes that diverge from fairness or from equilibrium, as we believe most nodes will do so in an attempt to get an advantage over the other nodes. The wolf-pack strategy will try to always find the most successful node or nodes and penalize them. We believe that just like in nature, a number of small predators can take down the bigger, more profitable ones. Furthermore during the Challenge we test two different strategies that provide completely opposite results. These offer a clear picture of what the best strategy is and the problems of the current system.

*Keywords— game theory, trust, ad-hoc networks, fairness.*

## I. Introduction

The MANIAC Challenge raises a number of interesting problems for ad-hoc networks: Can an ad-hoc network be a viable alternative for the currently used infrastructure-based networking? If infrastructure owners will use an ad-hoc network to off-load some of its traffic to have a more efficient overall network, how will the nodes in the ad-hoc network respond? What is a fair payment method for the nodes in the ad-hoc network? What are the risks that some nodes will abuse this new network to maximize its own profit?

In such a game, with every node trying to maximize its own profit the used strategies will intertwine making the end result extremely difficult to predict.

Based on previous competitions and the rules offered the following assumptions can be made: Most nodes, if not all will act in a rational way, trying to maximize its individual profit while still playing in a fair way so that they will not be eliminated from the network; Most nodes will try to stay as close as possible to the Nash Equilibrium [1, 2] while still trying to manipulate the numbers in their own favor; The entire network topology is known at all times through the OLSR protocol[4]; At all times at least one node will be connected with the wired backbone; Nodes will want close to complete fairness both in the rewards offered and the risk taken in the form of fines.

The idea of using mobile nodes to decongest the main infrastructure is clearly an important path to consider for the future of our networks and has been explored in various articles like [5]. Observing how multiple strategies work in such an environment is the first step in this direction. The possibility that one could abuse the network and create even worse performance needs to further be explored through contests of this type.

## II. Fairness in the Ad-Hoc Network

We strongly believe that all nodes participating in forwarding will act in a rational manner based on the Nash Equilibrium [1, 2] and as such the nodes in the network will behave fairly.

We define fairness [3] in the case of the MANIAC Challenge as the state when all nodes from source to the destination will equally split the profit and the fine needed to be paid if the packet is not delivered. In the case of bidding fairness would be achieved by always letting the node that is part of the shortest path win. As such the overall profit of all the nodes in the network will be maximized, extracting the largest possible amount from the back-bone infrastructure. Small deviations from fairness are accepted as this would permit differentiation in the bidding process.

If all nodes are expected to act fair it is easy to observe that the nodes that are unfair, or bad, will be penalized both through packet drop and forcing them to pay part of the fine or through refusal to let them win bids, even if their offer is a lot better than others.

In our proposed strategy we take advantage of this behavior by assisting bad nodes and penalizing rich nodes.

## III. The Wolf-pack strategy

We propose the Wolf-pack strategy, a novel way to approach the MANIAC Challenge. In this strategy we observe 2 key features and make decisions based on them: the network topology and the individual node behavior based on the network topology.

The network topology can be obtained through OLSR and it is available at any moment. It is important to be aware of the network topology, as it directly affects the reward each node should receive if the transactions are completely fair. By taking the node topology into consideration certain advantageous positions can be recognized and exploited, such as being the only node that can bid for a packet and thus modifying the strategy accordingly.

Individual node behavior is extremely important for our strategy. We need to make assumptions about how rich a node is and we do this by observing all transactions and calculating its profit at every point. We also need to observe how each node behaves, since a "bad" node can prove extremely useful. Information about individual nodes should be exchanged between our 2 friendly nodes to get a more accurate image of the entire network. We presume we can make our observations by packet sniffing but a solution in the case this is not possible is offered in section VI.

We believe that bad or unfair nodes will be penalized by all other nodes and as such their chances to win the challenge are greatly diminished. This nodes do not prove to be a direct challenge for us as such we will help them: if such a node bids for our packet and does not have an extreme tendency to drop packets discovered in the past we will prefer it as the next hop; If it has a package and wants to forward it, we will offer bids that give him a larger reward then and thus winning the bid and receiving the packet. As such we will increase our own number of delivered packages as we will take advantage of nodes other would not consider and we make sure the distribution of reward is spread across more nodes. This helps us in assuring that no node will collect a large number of rewards.

Next we will penalize the rich nodes. By observing the network we decide which are the nodes that manage to make a big profit so far and we chose other nodes to forward our packets. This is where the wolf-pack name comes in, we share the value of the profit we believe all other nodes have and directly attack the nodes with the most profit by not choosing them as forwarders and by dropping packets in an attempt to force them to pay big fines.

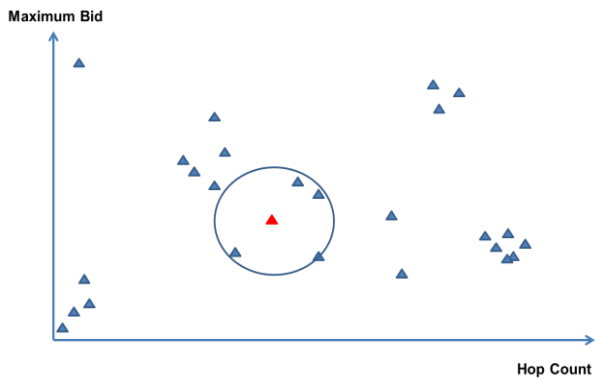

Fig 1 Bids arranged by Hop count and Maximum allowed bid matching them

By applying the two rules we prefer bad and poor nodes over rich and fair nodes in both bidding and forwarding, in the case of bidding we can go to the extra length of making a good bid and dropping the packet just to force the rich node to pay part of the fines. Based on this observations we order the bids we get based on the price offers we receive, the level of profit the node has and the node behavior from the past and we choose the most advantageous for us to disseminate the total network profit as equally over as many nodes as possible while keeping ourselves just a bit more above anyone so we do not become a clear target for other nodes.

IV. IMPLEMENTATION

To implement our strategy we require the four following metrics:

- Node richness – this is observed over time by listening to bid win messages and determining the profit for every node.
- Node Fairness – it is computed by determining if the amount kept by a node is the expected one or if it deviates.
- Topology – is provided by the OLSR protocol.
- Bid data – provided per individual.

Out of the 4 metrics only the last one is considered to always be accurate. The Topology changes at a very high rate and we can't expect it to be up to date at every point, furthermore we are only provided with the 1 hop and 2 hop neighbors. The fairness and richness of individual nodes can only be accurate if we receive all the messages in the network; this is not the case, we only receive the 1 hop messages.

For each of this metrics a sorted list is created containing every node that made a bid for our current packet. We prefer small values for richness, topology distance to the destination using the given node, bid offered by the node and its fairness. This matches what we described in the previous section. These 4 lists are then combined using a weight for each list and the best node is chosen as the next one for our packet.

When we make a bid we try to predict what the other nodes will bid and bid less. However we don't want to make bids that are too low. We use historical data of the network and when we receive a bid request we make the smallest bid compared to the past bids made by the other participating nodes taking into account the similarity with the previous bids.

To achieve this we only choose the past bids that are closer than a distance ε to our current scenario. The distance is calculated by sorting all the bids in a graph according to Maximum allowed Bid and Hop count to the current bid as can be observed in *Fig. 1*.

This method is especially important considering during the contest we noticed a lot of bids with the value 0. This has led us to believe that some algorithms chose the smallest bid possible at all times and converged on 0. The method we used guarantees that as long as there are not too many bids of 0 our strategy would still make useful bids while bidding the smallest one if possible.

In a real life scenario the algorithm can be extended by only keeping a history of the last N number of bids we saw or by limiting the history to the last x minutes.

## V. CHANGES DURING THE CONTEST

In the first round our algorithm managed to rank third. Because we wanted to get to the first position we decided to make dramatic changes to guarantee that we make a profit. We tried two additional strategies.

One was to always bid 1 and keep the part of the algorithm that dealt with who receives a packet that we already have. We expected this to work because we observed many bids of 0 and we hoped that we could deprecate the network fast enough that only we would make profit whenever we managed to win a packet. This strategy further guarantees that we would receive a large number of packets at the start.

The strategy ranked last. This happened because of the large fines; the large fines guaranteed that if one bids too low he will lose a lot of points.

Our next and final strategy was based on the observations we made in the previous rounds. First we wanted to keep the bids high. Next we wanted to get only the messages that are guaranteed to be delivered, thus avoiding any fines. To achieve this meant bidding according to the Wolf-pack strategy, but only when we are 1 hop away from the destination. A variation could have been to only bid a value of 1.

Because of the way the API was build one had to always bid, even when this was not wanted, first we tried to beat this system by always bidding max bid, but even like this the fines were too large and we did not manage to make a profit. Then we observed a problem we had in the first version.

Because of the way we calculated our bids and the large amount of time it took to do so we were missing a lot of bids. First we saw this as a problem and it resulted in the second version where we always bid 1. But now with a clear picture we saw it as an advantage. We knew at this point that the only guaranteed way to make a profit was to only bid when we were next to the destination. When this was not the case we wanted not to play, we wanted to delay the bid as much as possible. The problem introduced by the first algorithm with the big delay was now an advantage. We knew how to delay the messages, how to reliably participate in the bidding process only when it brought us a profit.

We delayed the bid as much as possible and then started a large processing to assure that every bid we didn't want to make would time out. The bids that were guaranteed to give us profit were executed as fast as possible by keeping a reduced history or by randomly bidding a very small value. At this point we could only make profit. This final strategy threw us in the first place for the last round.

## VI. DISCUSSION

In the case we cannot sniff the network to obtain data of what bids other nodes have made or who won a bid we propose the following strategy: We keep a history of all bids; when we find ourselves in a bid against the same other nodes we apply a binary search method. We bid the center between the lowest possible bid and the highest, if we win the next time we bid the center between the highest possible bid and the previous center; if we do not win we bid the center between our bid and the possible lowest. We continue to bid as such until we find the bid each other node makes and the strategy they play. We compare each bid with the strategy we believe the other nodes can play in an attempt to find the strategy and the level of fairness they are at.

During the contest we were able to obtain all the data and this was not required.

## VII. CONCLUSION

In this paper we presented the wolf-pack strategy. This strategy aims to win the MANIAC Challenge by attacking the rich nodes and helping the poor and even the bad nodes in the network. During the contest we observed that the initial strategy was successful only in a small majority of cases obtaining a rank of 3. We implemented and executed 2 new strategies during the contest. These 2 strategies obtained us the last and first place and gave us an important insight into the entire goal of the MANIAC Challenge, off-loading data to the nodes. Even though the last strategy shows that the system can indeed be tricked, we believe that with small modifications to the rules data off-loading can provide a future of congestion free networks. From our experience more study in the way the fine affects the network or how the Time-To-Live limits the number of successfully delivered messages is required. We believe different methods of applying the fine or by removing it all together could bring completely different results.